\pdfoutput=1
\documentclass[svgnames]{article}

\usepackage[nonatbib, final]{neurips_2024}
\usepackage[sort, numbers, compress]{natbib}

\usepackage{xcolor}
\usepackage{microtype}
\usepackage{hyperref}
\usepackage{url}
\usepackage{enumitem}
\usepackage{booktabs}
\usepackage{graphicx}
\usepackage{color-edits}
\usepackage{booktabs, multirow}
\usepackage{pdflscape}
\usepackage{amsmath, amsfonts, amssymb}
\usepackage{bbm}
\usepackage{nicefrac}
\usepackage{microtype}

\hypersetup{colorlinks=true, citecolor=Navy, linkcolor=blue, urlcolor=blue}

\title{A Shared Standard for Valid Measurement\\ of Generative AI Systems'\\ Capabilities, Risks, and Impacts}

\author{%
    Alexandra Chouldechova\thanks{Corresponding author: \texttt{alexandrac@microsoft.com}} \ \ \ 
    Chad Atalla \ \ \ 
    Solon Barocas \ \ \ 
    A. Feder Cooper \ \ \ 
    Emily Corvi \\
    \textbf{
    P. Alex Dow \ \ \ 
    Jean Garcia-Gathright \ \ \ 
    Nicholas Pangakis \ \ \ 
    Stefanie Reed}  \\
    \textbf{Emily Sheng \ \ \ 
    Dan Vann \ \ \ 
    Matthew Vogel \ \ \ 
    Hannah Washington \ \ \ 
    Hanna Wallach}    \\
  Microsoft Research
}

\begin{document}

\maketitle

\vspace{-.5cm}
\begin{abstract}
The valid measurement of generative AI (GenAI) systems' capabilities, risks, and impacts forms the bedrock of our  ability to evaluate these systems. 
We introduce a shared standard for valid measurement that helps place many of the disparate-seeming evaluation practices in use today on a common footing.  Our framework, grounded in measurement theory from the social sciences, extends the work of \citet{adcock2001measurement} in which the authors formalized \textit{valid measurement} of concepts in political science via three processes: \emph{systematizing} background concepts, \emph{operationalizing} systematized concepts via annotation procedures, and \emph{applying} those procedures to instances.  We argue that valid measurement of GenAI systems' capabilities, risks, and impacts, further requires systematizing, operationalizing, and applying not only the entailed \textit{concepts}, but also the \textit{contexts} of interest and the \textit{metrics} used.  This involves both \textit{descriptive} reasoning about particular instances and \textit{inferential} reasoning about underlying populations, which is the purview of statistics. By placing many disparate-seeming GenAI evaluation practices on a common footing, our framework enables individual evaluations to be better understood, interrogated for reliability and validity, and meaningfully compared.  This is an important step in advancing GenAI evaluation practices toward more formalized and theoretically grounded processes---i.e., toward  a science of GenAI evaluations.\looseness=-1 
\end{abstract}

\vspace{-.3cm}
\section{Introduction}

Whether we are interested in what a generative (GenAI) system is capable of, the risks it poses, or the impacts of its deployment, we can gain insight into such inquiries by framing them as \textit{measurement tasks} of the form: \emph{measure the [amount] of a [concept] in [instances] from a [population]}. Examples include evaluating an LLM-based tutoring system's mathematical reasoning capabilities by measuring its average (amount) performance (concept) on standardized test questions (instances) from the Korean CSAT Math Exam (population); or evaluating 
the risks posed by a conversational search system by measuring the prevalence (amount) of stereotyping (concept) in the system's outputs (instances) in the current US deployment (population).
This template is also sufficiently expressive to capture many of the 
impact 
evaluation tasks described by 
\citet{solaiman2023evaluating}. 
Recent work has emphasized the need, when carrying out such measurement tasks, to precisely articulate \textit{what} is being measured---i.e., to formalize the \textit{concept} of interest.  This includes critical work on dataset diversity \citep{zhao2024position}, benchmark design for stereotyping and other concepts \citep{blodgett2021stereotyping, EBCD}, and model memorization \cite{carlini2022quantifying, ippolito2023preventingverbatimmemorizationlanguage, cooper2024files}.\looseness=-1 

These critiques parallel those made two decades ago by \citet{adcock2001measurement}, who argued that political scientists were devoting insufficient attention to measurement validity---i.e., the question of whether researchers' reported measurements adequately reflect the concepts they purport to measure.  To facilitate this type of reflection, Adcock \& Collier introduced a four-level framework that formalizes 
the relationship between a concept of interest and reported measurements for that concept as a structured progression from a \textit{background concept} (the set of meanings and understandings associated with the concept), to a \textit{systematized concept} (precise definitions), to \textit{annotation procedures} (procedures for labeling or scoring instances) that operationalize the systematized concept, to the \textit{application} of those procedures to obtain scores or labels for one or more instances. Validity concerns about concepts then arise as slippage between these four levels, such as a failure of the annotation procedures to capture relevant dimensions of the systematized concept. Using this framework, the above-mentioned critiques of GenAI evaluations can  be posed as failures to systematize nebulous background concepts (e.g., stereotyping) before jumping to operationalization via annotation procedures (e.g., instructions that ask crowdworkers to label system outputs as stereotyping or not).\looseness=-1

\vspace{-.05cm}
\section{Framework Overview}
\vspace{-.05cm}

Although we wholeheartedly agree with these critiques, we argue that equal attention should be paid to validity concerns that arise from the under-specification of \textit{amounts}, \textit{populations}, and \textit{instances} entailed in such measurement tasks. 
In other words, valid measurement of GenAI systems' capabilities, risks, and impacts requires systematizing, operationalizing, and applying not only \textit{concepts}, but also \textit{amounts}, \textit{populations}, and \textit{instances}. This involves both \textit{descriptive} reasoning about particular instances and \textit{inferential} reasoning about underlying populations, which is the purview of statistics. 

\begin{figure}[!t]
    \centering    \includegraphics[width=1\linewidth]{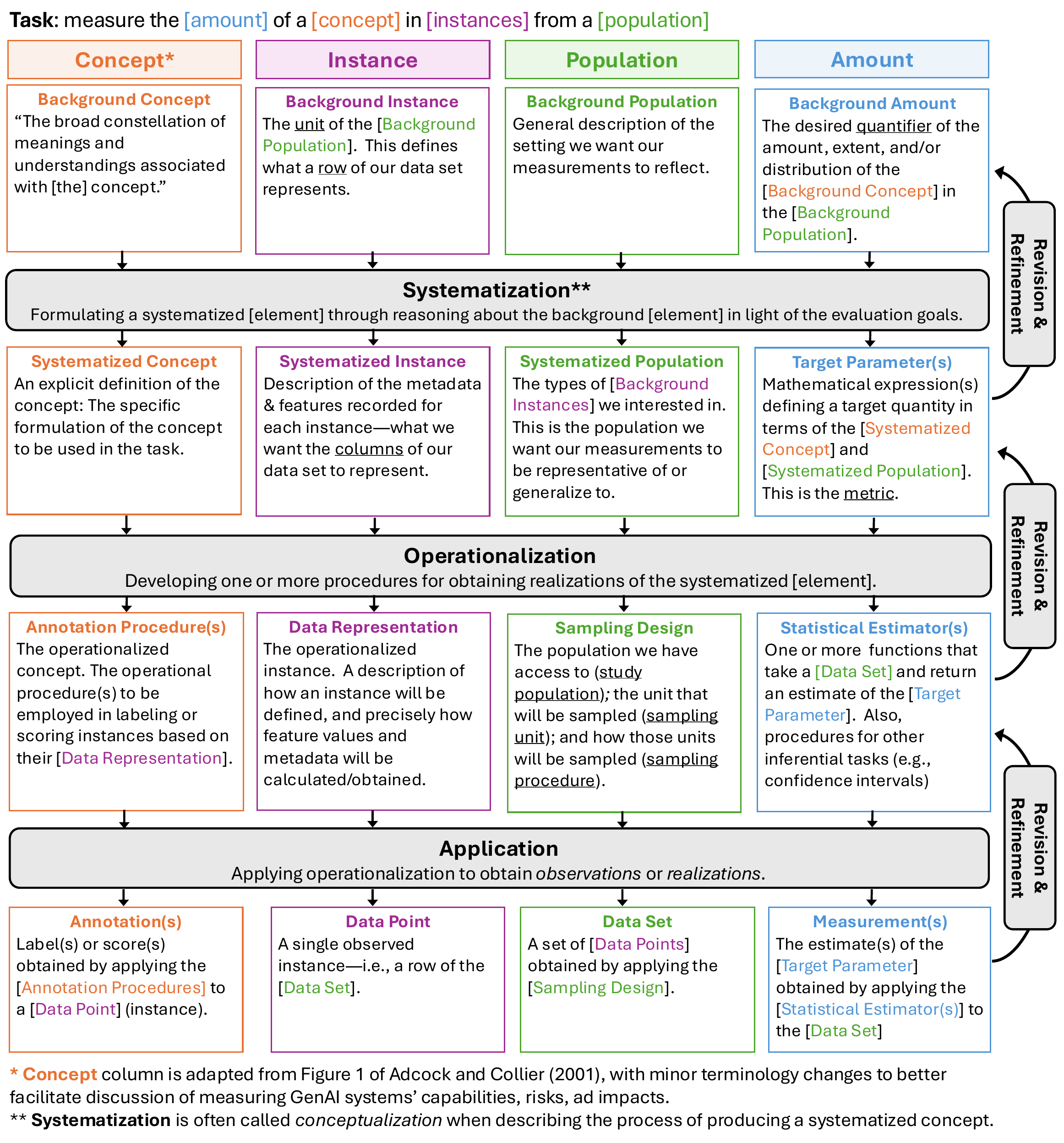}
    \vspace{-.4cm}
    \caption{Our proposed framework for measurement tasks of the form: \textit{measure the [amount] of a [concept] in [instances] from a [population]}.  The figure shows how the four elements that make up such tasks---amounts, concepts, instances, and populations---are formalized through the sequential processes of \textit{systematization}, \textit{operationalization} and \textit{application}.  Elements in earlier levels (rows)  can be revised and refined based on findings, including  validity concerns, that arise in later levels.}
    \vspace{-.2cm}
    \label{fig:4by4_ttable}
\end{figure}

To facilitate reflection on such validity concerns, we introduce an extension of Adcock and Collier's framework~\citep{adcock2001measurement}.  A graphical representation of this framework is depicted in Figure~\ref{fig:4by4_ttable}. We developed and iteratively refined the framework using numerous examples of GenAI and non-generative AI systems' capabilities, risks, and impacts, including topic modeling, representational harm measurement for multi-modal models, face verification technologies, automated speech recognition, and others.\looseness=-1

Like Adcock \& Collier's framework, our framework includes processes, depicted as backward arrows, for revising and refining earlier levels based on findings, including validity concerns, that arise in later levels. These revision and refinement processes operate not only within each column (e.g., within the ``Concept'' column, which is adapted from Adcock \& Collier's framework), but also across the columns.  We may find, for example, that the variance of our measurements is too high, leading us to revise the sampling design or even the annotation procedures to improve estimator stability.\looseness=-1

\begin{figure}[!t]
    \centering    \includegraphics[width=1\linewidth]{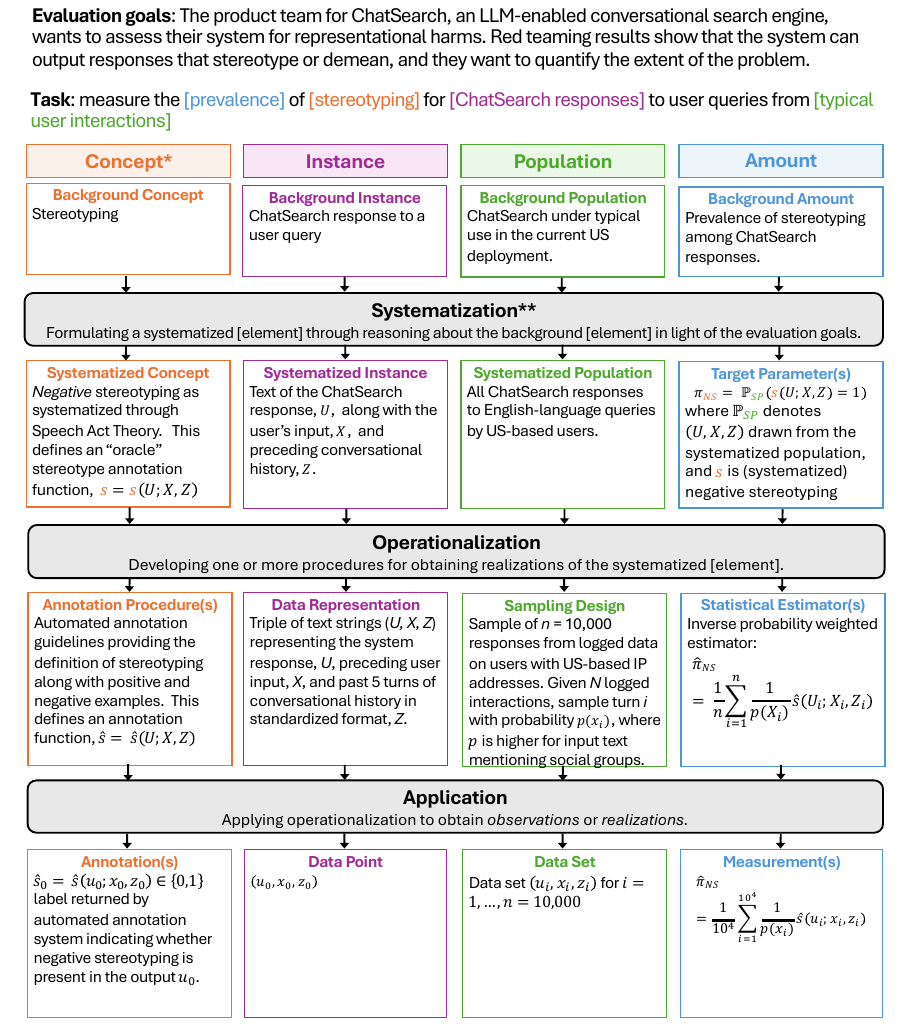}
    \caption{A high-level example of using our framework in a hypothetical evaluation of a conversational search engine, ChatSearch.  Each cell provides an overview of what a complete measurement procedure instantiated using the framework could look like. Note that a full instantiation would require providing considerable additional information.  For example, a fully systematized complex concept or the full description of a complex sampling design might require several pages of exposition.}
    \label{fig:chatsearch}
\end{figure}

To better understand the elements of our framework, consider the example of measuring stereotyping in a conversational search system. How we represent an instance (e.g., as a system output vs. as a system output and its corresponding input) may influence whether a system output is labeled as stereotyping. Likewise, if we are interested in the prevalence of stereotyping under typical use post-deployment, but our sampling design (operationalized population) is based on adversarial red teaming, we are likely to vastly overestimate the prevalence. By structuring the choice of metric as first defining a \textit{target parameter} (systematized amount) that is then estimated using statistical estimators (operationalized amount), we can apply established statistical inference methods to optimally trade off between bias and variance, adjust for sample bias in the data set (i.e., correct for mismatches between the systematized population and the sampling design), construct uncertainty intervals, and perform hypothesis tests.  In Figure~\ref{fig:chatsearch}, we provide a high-level example of using our~framework to measure stereotyping in a hypothetical conversational search engine, ChatSearch.\looseness=-1

\vspace{-.05cm}
\section{Limitations and Future Work}
\vspace{-.05cm}

Although our framework brings structure to the question of \textit{what} should be done when measuring GenAI systems' capabilities, risks, and impacts, we offer no guidance on \textit{how} to accomplish such measurement tasks. This limitation therefore remains an important direction to explore in future work.\looseness=-1

A major limitation of the framework itself is that it does not help with formulating measurement tasks in the first place, nor does it offer guidance on how to interpret resulting measurements or what decisions those measurements can or should inform.  Just as the initial problem formulation can have a big effect on the fairness of an ML model \cite{laufer2023optimization, obermeyer2019dissecting, passi2019problem,  watson2023multi, qian2021are} or the conclusions drawn about deep-learning optimizer performance~\cite{dodge2019nlp, cooper2021deception, obandoceron2024consistencyhyperparameterselectionvaluebased}, what we choose to measure, in what population, and how we quantify it matters a great deal \cite{coston2020counterfactual, mulvin2021proxies, cooper2024variance}. The revision and refinement processes provide one mechanism for revisiting these decisions, but this falls well short of addressing the full complexity of formulating~measurement tasks---an issue that is critically important to evaluations of GenAI systems.\looseness=-1

\section{Conclusion}

Our framework helps place many of the disparate-seeming GenAI evaluation practices in use today on a common footing. This enables individual evaluations to be better understood, interrogated for reliability and validity, and meaningfully compared. In this way, the framework is intended to serve as a shared standard for valid measurement of GenAI systems' capabilities, risks, and impacts. For example, using it allows us to (1) more easily identify and remedy validity concerns; and (2) compare different measurement tasks by identifying precisely where within the framework they diverge. In other words, our framework enables us not only to ``evaluate evaluations'' but also to improve how evaluations of GenAI systems are designed in the first place, thereby helping advance such evaluations from their current state of disparate-seeming and ad hoc practices~\citep{roose2024ai, maslej2024ai} toward more formalized and theoretically grounded processes---i.e., toward a science of GenAI evaluations.\looseness=-1

\small
\bibliography{references}
\bibliographystyle{plainnat}

\end{document}